\documentclass[final,1p,times]{elsarticle}
\usepackage{amssymb}
\usepackage{amsmath}
\usepackage{lineno}
\usepackage{hyperref}
\usepackage{xcolor}
\begin{document}
%\linenumbers
\begin{frontmatter}

%% Title, authors and addresses

%% use the tnoteref command within \title for footnotes;
%% use the tnotetext command for theassociated footnote;
%% use the fnref command within \author or \affiliation for footnotes;
%% use the fntext command for theassociated footnote;
%% use the corref command within \author for corresponding author footnotes;
%% use the cortext command for theassociated footnote;
%% use the ead command for the email address,
%% and the form \ead[url] for the home page:
%% \title{Title\tnoteref{label1}}
%% \tnotetext[label1]{}
%% \author{Name\corref{cor1}\fnref{label2}}
%% \ead{email address}
%% \ead[url]{home page}
%% \fntext[label2]{}
%% \cortext[cor1]{}
%% \affiliation{organization={},
%%             addressline={},
%%             city={},
%%             postcode={},
%%             state={},
%%             country={}}
%% \fntext[label3]{}

\title{AURORA: A High Performance Software DAQ Framework for Next-Generation Rare-Event Search Experiments}

%% use optional labels to link authors explicitly to addresses:
%% \author[label1,label2]{}
%% \affiliation[label1]{organization={},
%%             addressline={},
%%             city={},
%%             postcode={},
%%             state={},
%%             country={}}
%%
%% \affiliation[label2]{organization={},
%%             addressline={},
%%             city={},
%%             postcode={},
%%             state={},
%%             country={}}

\author[SJTU_PHY]{Yihan Guo}
\author[SJTU_PHY]{Xiaofeng Shang}
\author[THU]{Chang Cai}
\author[SJTU_PHY]{Weihao Wu}
\author[TDLI,SJTU_NEW,SJTU_SC]{Xun Chen\corref{cor}}
\ead{chenxun@sjtu.edu.cn}

\cortext[cor]{Corresponding author}
%% Author affiliation
\affiliation[TDLI]{organization={State Key Laboratory of Dark Matter Physics, Tsung-Dao Lee Institute, Shanghai Jiao Tong University},%Department and Organization
            city={Shanghai},
            postcode={201210}, 
            country={China}}
\affiliation[SJTU_PHY]{organization={School of Physics and Astronomy, Shanghai Jiao Tong University, Key Laboratory for Particle Astrophysics and Cosmology (MoE), Shanghai Key Laboratory for Particle Physics and Cosmology},city={Shanghai}, postcode={200240}, country={China}}
\affiliation[THU]{organization={Department of Physics, Tsinghua University}, city={Beijing}, postcode={100084}, country={China}}
\affiliation[SJTU_NEW]{organization={New Cornerstone Science Laboratory, Tsung-Dao Lee Institute, Shanghai Jiao Tong University}, city={Shanghai},
            postcode={201210}, 
            country={China}}
\affiliation[SJTU_SC]{organization={Shanghai Jiao Tong University Sichuan Research Institute}, city={Chengdu}, postcode={610213}, country={China}}

\begin{abstract}
%% Text of abstract
The upcoming PandaX-xT experiment will deploy over 3,000 readout channels operating at a 500~MSa/s sampling rate, generating a sustained data bandwidth up to 1.6~GB/s. To meet this demanding requirement, we present AURORA, a high-performance, distributed data acquisition (DAQ) framework designed for scalability, low latency, and efficient resource utilization. Built on a modular architecture and leveraging modern I/O and networking technologies, including multi-level buffering, deferred and asynchronous processing, AURORA achieves a projected throughput of over 3~GB/s on the aggregation node in benchmark tests. While developed to support PandaX-xT, the framework is experiment-agnostic and readily adaptable to other large-scale particle and nuclear physics experiments.
\end{abstract}

%% Keywords
\begin{keyword}
%% keywords here, in the form: keyword \sep keyword
xenon detector \sep DAQ \sep neutrino \sep dark matter
%% PACS codes here, in the form: \PACS code \sep code

%% MSC codes here, in the form: \MSC code \sep code
%% or \MSC[2008] code \sep code (2000 is the default)

\end{keyword}

\end{frontmatter}

%% Add \usepackage{lineno} before \begin{document} and uncomment 
%% following line to enable line numbers
%% \linenumbers

%% main text
%%

%% Use \section commands to start a section
\section{Introduction}
\label{sec:intro}
The PandaX experiment series~\cite{PandaX:2014mem, PandaX:2016pdl, PandaX:2018wtu} employs dual-phase xenon time-projection chambers~\cite{Aprile:2009dv} to investigate fundamental physics, including dark matter direct detection, solar neutrino observations, and the search for neutrinoless double-beta decay. 
The current running PandaX-4T experiment has already produced a number of significant scientific results~\cite{PandaX-4T:2021bab, PandaX:2022kwg, PandaX:2022aac, PandaX:2023toi, PandaX:2024muv, PandaX:2024qfu}.
The next-generation experiment, PandaX-xT, will feature a xenon target exceeding 40 tonnes, significantly advancing the sensitivity frontier for dark matter interactions~\cite{PANDA-X:2024dlo}.

According to the conceptual design report (CDR)~\cite{PANDA-X:2024dlo}, the final detector will be equipped with up to 8000 readout channels. The first phase of PandaX-xT, designated PandaX-20T, is planned to operate with around 3000 channels. Extrapolating from the PandaX-4T experience, where the sustained science data rate is about 0.1~MB/s per channel, the total aggregate data rate is projected to be 300~MB/s for the initial stage and 800~MB/s for the final configuration. During calibration runs with intensive internal or external sources, the rate may double, reaching up to 1.6~GB/s. An additional Outer Veto system will further contribute to the total data volume. All data must be acquired and stored reliably for offline analysis.

The current Data Acquisition (DAQ) system operating for PandaX-4T is a triggerless, distributed architecture~\cite{Yang:2021hnn}. It utilizes six acquisition servers, each reading data from eight digitizer boards via dedicated optical links. The data are then transmitted over a 10~Gbps network to a central event-builder server for assembly and storage. This system is designed for a maximum throughput of 800~MB/s and has operated stably at 300~MB/s during physics runs.

However, during prototype tests for PandaX-20T setup at the Tsung-Dao Lee Institute, Shanghai Jiao Tong University, the existing DAQ system was not able to sustain data rates of approximately 600 MB/s. This performance limitation highlights the urgent need for a more robust and scalable DAQ solution capable of meeting the demands of future, larger-scale detectors.

Several DAQ systems have been developed for contemporary rare-event-search experiments.  XENONnT employs a triggerless DAQ with FPGA-based zero-length encoding, sustaining readout rates of approximately 500~MB/s during calibration runs~\cite{XENON:2022vye}. The LZ experiment uses the FPGA-based FADR system, which performs pulse-only digitization to reduce raw waveform data by up to a factor of 50 before forwarding the selected data to distributed event builders~\cite{LZ:2024bvw}.  For the JUNO neutrino observatory, a distributed DAQ framework handles over 40~GB/s of raw data, but after online event compression the volume recorded to storage is approximately 60~MB/s~\cite{Chen:2025phw}.  While each of these systems is well tuned to its own experimental requirements, their design choices differ from the needs of PandaX-xT, where the full raw waveform must be preserved without online reduction at a high aggregate rate exceeding 1~GB/s.

To address these challenges, we have developed the Adaptable Unified Real-time Online Readout Architecture (AURORA), a new high-performance DAQ framework designed for the PandaX-xT experiment and beyond. The framework employs a multi-level buffering strategy and deferred processing techniques to achieve a maximum throughput of up to 3~GB/s. This article details the design, implementation, and performance evaluation of the AURORA framework. Section~\ref{sec:requirements} outlines the system requirements and design goals. The overall architecture is described in Section~\ref{sec:architecture}. Performance benchmarks and stability tests are presented in Section~\ref{sec:performance}. The scalability and general applicability of the framework are discussed in Section~\ref{sec:scalability}. Finally, a summary and outlook are provided in Section~\ref{sec:summary}.
\section{System Requirements}
\label{sec:requirements}

The design of the data acquisition system is driven by the stringent performance and operational demands of the next-generation PandaX experiment. The core requirements are categorized as follows.

\subsection{Performance and Scalability}

The system must handle the high data volume from nearly 3000 readout channels. A primary requirement is the capability to reliably maintain a continuous data throughput of 1.0 GB/s during normal operation. This requirement, while far above the ~300 MB/s expected for 3000 channels under standard underground conditions, is necessary to guarantee stable performance during ground testing, where much higher background levels drastically increase the data volume. Furthermore, the architecture must be designed to process short-term data bursts exceeding 1.6 GB/s without loss, ensuring that it can cope with peak activity periods. To physically and logically accommodate the large number of channels, the system must support a distributed architecture composed of at least 8 independent data acquisition servers, each running an instance of the data-taking process (\textit{daq\_reader}). The communication framework between the central data collection program (\textit{collector}) and these servers must be highly efficient for data transfer and easily extensible to support future expansion in the number of servers or channels.

\subsection{Data Integrity and Reliability}

Guaranteeing the completeness and correctness of the physics data is paramount. The system must ensure zero data loss from the point of packet reception at the \textit{daq\_reader} to the point of persistent storage on disk. This requirement necessitates robust end-to-end error handling, reliable delivery mechanisms, and careful management of system buffers. In addition to data integrity, overall system stability is critical. The framework must demonstrate high intrinsic stability and be capable of continuous operation for periods of no less than 24 hours under conditions of normal hardware functionality and without requiring manual intervention.

\subsection{Multi-Stream Data Handling}

Beyond the primary time projection chamber (TPC), the experiment requires the integration of data from an external veto system for background rejection and in-situ measurement of atmospheric neutrino events.
To accommodate this, the DAQ system is designed around the concept of logical data streams. Each subsystem is assigned a dedicated data stream (e.g., TPC Stream and VETO Stream). A single stream may be serviced by one or more physically distinct data acquisition servers, all producing data packets with an identical format. The core requirement for the \textit{collector} is to support the concurrent receiving and processing of multiple, independent data streams. It must maintain separate buffering and sorting queues per stream to preserve internal data ordering, and finally persist the data of each stream to segregated storage. This logical separation at the acquisition stage is crucial for enabling efficient, independent offline analysis of each subsystem's data, while the common timing framework allows for correlated analysis across streams when required.

\section{Architecture Design}
\label{sec:architecture}

\begin{figure}[htb]
    \centering
    \includegraphics[width=0.95\textwidth]{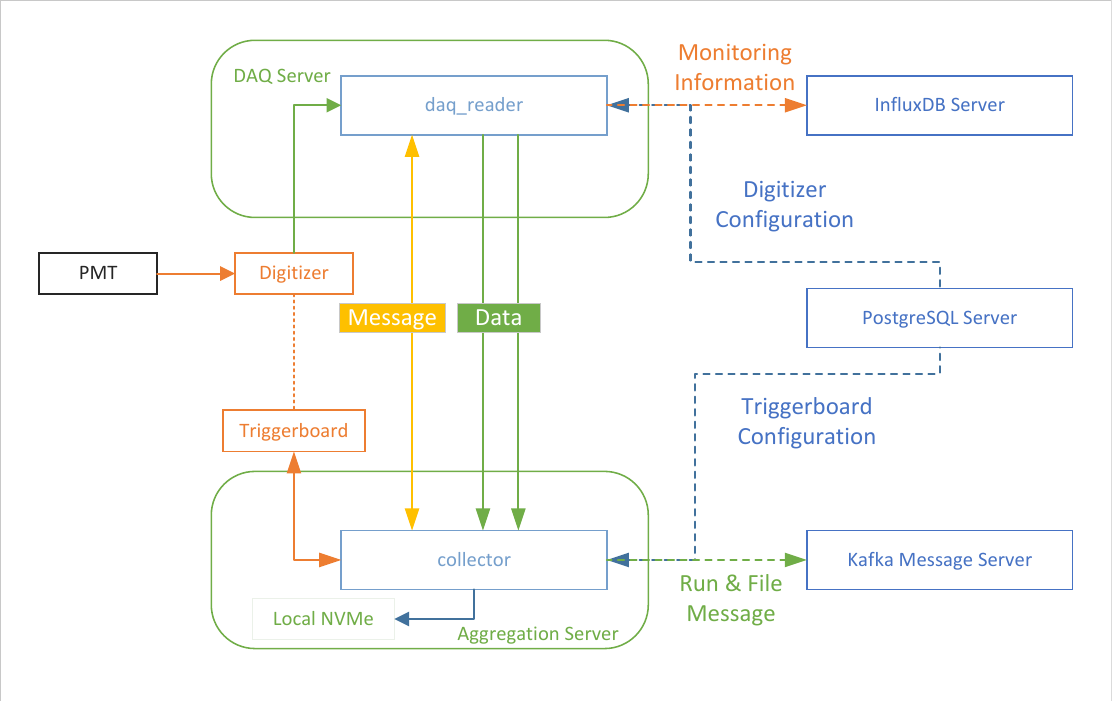}
    \caption{Schematic overview of the distributed DAQ architecture. For clarity, only one representative PMT, digitizer, and DAQ server are shown, illustrating the data flow from the detector frontend to the aggregation server. Additionally, the diagram indicates the system’s interaction with key external services: configuration is retrieved from a PostgreSQL server; monitoring metrics are sent to an InfluxDB server; and run and file information are published to a Kafka message server.}
    \label{fig:architecture}
\end{figure}

To address the requirements outlined in the previous section, particularly those concerning high throughput, scalability, multi-stream handling, and data integrity, we have designed a distributed DAQ system whose overall architecture is illustrated in Fig.~\ref{fig:architecture}.

At the frontend, each digitizer converts analog voltage signals from 8 photomultiplier tubes (PMTs) into digital data~\cite{He:2021sbc}. These digitizers communicate with the DAQ servers via optical fibers using the hardware-based TCP processor (SiTCP) ~\cite{uchida2008hardware}. Each DAQ server is equipped with 7 network adapters, each providing 4 SFP ports, allowing a single server to connect to up to 28 digitizers. The \textit{daq\_reader} program runs on each DAQ server, continuously reading out data from the attached digitizers during data taking.

A dedicated aggregation server is linked to each DAQ server via a direct 10 Gbps SFP+ connection. Running on this server, the collector program receives the incoming data streams, identified by their source IP addresses. Packets belonging to the same stream are sorted based on embedded timestamps before being written to a local high-speed NVMe SSD. This design ensures both high throughput and reliable, time-ordered storage for subsequent offline analysis. The aggregation server is also connected to the trigger board, which manages the global hardware clock and controls the data-taking lifecycle of all digitizers. The \textit{collector} can send control commands to the trigger board and retrieve hardware clock information. 

External services provide essential operational functions for the framework. Operational metrics from \textit{daq\_reader}, such as real-time data bandwidth, can be sent directly to a time-series database (InfluxDB~\cite{influxdb}) if the service is available. Detailed configuration parameters for the digitizers and trigger board, including thresholds and trigger logic, are stored in a PostgreSQL database~\cite{postgresql}. Both \textit{daq\_reader} and \textit{collector} connect to this database to load the necessary configurations during the initialization of each run. An automatic data processing pipeline is orchestrated using the Kafka messaging service~\cite{Kreps2011KafkaA, Sax2018, Apache:Kafka}. Whenever the collector completes writing a data file, it publishes the corresponding run and file metadata to a dedicated ``daq'' topic in Kafka. This mechanism enables prompt downstream processing and transfer of the acquired data. 

\subsection{Raw data format}
To comprehend the data transfer mechanisms and the design of the multi-level buffering system, it is first essential to understand the raw data format produced by the digitizer.

\begin{figure}[htb]
    \centering
    \includegraphics[width=.95\textwidth]{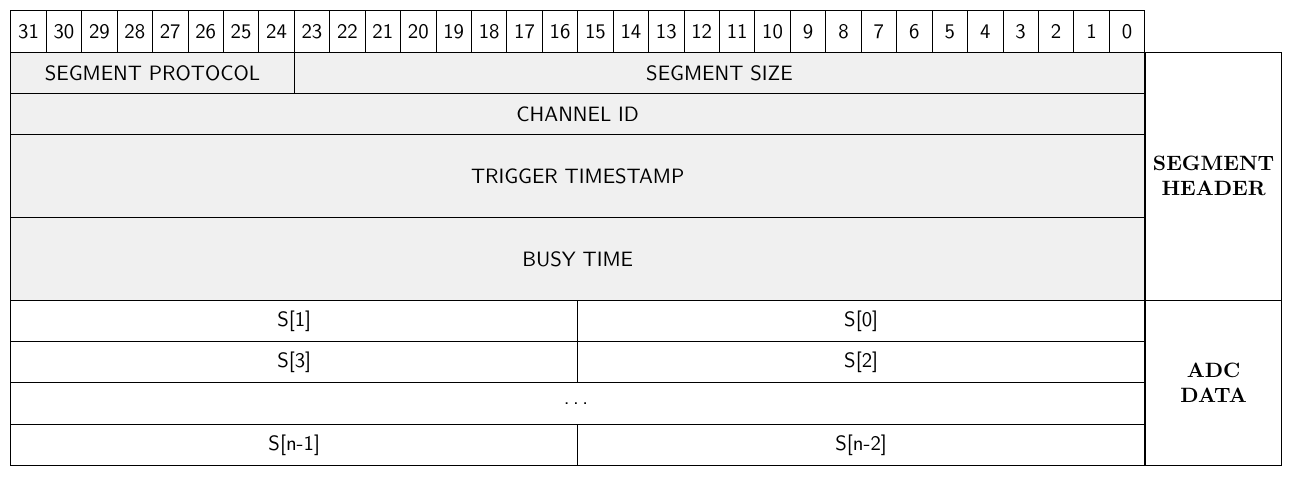}
    \caption{Structure of the raw data block.}
    \label{fig:data_block}
\end{figure}

The fundamental unit of raw data is called a block, whose structure is stored in \textit{little-endian} byte order, and shown in Fig.~\ref{fig:data_block}. A block corresponds to a continuously recorded waveform segment from a single readout channel. Each data block consists of a header followed by the ADC data. The header contains the following fields:
\begin{itemize}
    \item An 8-bit segment protocol (reserved as \texttt{0xb2} in current PandaX implementation),
    \item A 24-bit segment size specifying the total block size $N$ in bytes,
    \item A 32-bit channel number,
    \item A 64-bit timestamp indicating the segment's start time, and
    \item A 64-bit busy time (used solely for debugging).
\end{itemize}
Following the header, the ADC data are arranged sequentially. Each ADC sample is represented by a 16-bit (two-byte) unsigned integer.
Given the total block size $N$ (in bytes) specified in the header, the ADC data payload occupies $N-24$ bytes. Consequently, the number of samples $n$ in a block is calculated as $n = (N-24)/2$. The maximum number of continuous samples in a segment recorded by the digitizer is 65,534.

Each successful read operation via the digitizer library produces a data packet containing an integer number of complete data blocks. Blocks from all channels within the same data stream must be stored in chronological order, sorted by their embedded timestamp. Thus, the \textit{daq\_reader} must transfer data packets to the \textit{collector} without loss, while the \textit{collector} is responsible for sorting all incoming data blocks and writing them to storage.

\subsection{Network layer implementation}
Safe and high-throughput network transfer is essential for data delivery. This project employs the standalone Asio library\cite{boost_asio_docs}, a header-only C++ library designed for high-performance, low-level I/O programming.

The \textit{daq\_reader} program acts as a TCP server, while the \textit{collector} program operates as a TCP client. During data acquisition operations, multiple connections are established from the \textit{collector} client to each server. One connection is reserved for control commands and response messages, while the remaining connections are dedicated to unidirectional data transmission. This configuration enables efficient utilization of the available bandwidth.

The \textit{collector} controls all \textit{daq\_reader} instances through the \textit{control connections} using a simple, custom application protocol. The commands sent from the \textit{collector} to a \textit{daq\_reader} in this protocol are defined as follows:
\begin{itemize}
    \item \texttt{INIT}: Upon receipt, the \textit{daq\_reader} retrieves the necessary configuration from the database and initializes its digitizers. Once initialization is complete, the server responds with \texttt{READY}.
    \item \texttt{START}: Upon receipt, the \textit{daq\_reader} starts data acquisition threads for all managed digitizers and responds with \texttt{STARTED}.
    \item \texttt{STOP}: Upon receipt, the \textit{daq\_reader} program terminates all acquisition processes and returns a \texttt{STOPPED}. All connections to the server are shut down by the \textit{collector} upon receiving the response.
\end{itemize}
Acquired data are transmitted from the \textit{daq\_reader} to the \textit{collector} via the dedicated data connections. The transmission follows a simple packet format: \texttt{DATA N\textbackslash{}n[payload]}. Here, \texttt{N} is the number of bytes in the subsequent payload, represented as a decimal string. The payload itself is a data packet comprising one or more data blocks.

\begin{figure}[tbp]
    \centering
    \includegraphics[width=0.9\textwidth]{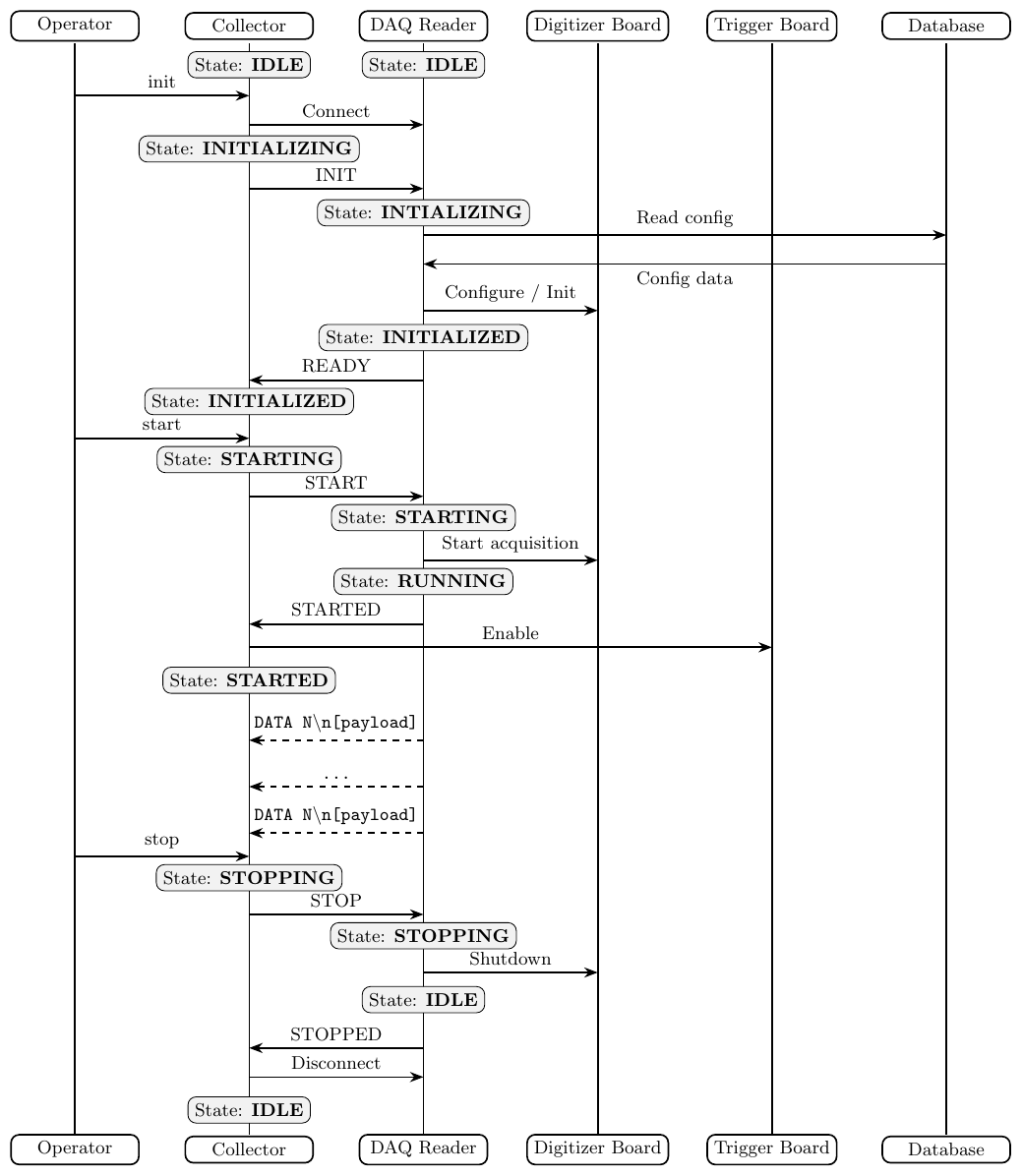}
    \caption{The flow of a data acquisition cycle and command history. The state change of the collector is shown.}
    \label{fig:daq_flow}
\end{figure}

Both the \textit{collector} and each \textit{daq\_reader} operate as explicit state machines, ensuring robust and orderly control of the distributed system. A typical acquisition cycle, illustrated in Fig.~\ref{fig:daq_flow}, is driven by the \textit{collector} in response to operator commands. The \textit{collector} progresses through states (\texttt{IDLE}, \texttt{INITIALIZING}, \texttt{INITIALIZED}, \texttt{STARTING}, \texttt{STARTED}, \texttt{STOPPING}) by issuing protocol commands (\texttt{INIT}, \texttt{START}, \texttt{STOP}) only when permissible, and advances only after receiving the expected acknowledgments (\texttt{READY}, \texttt{STARTED}, \texttt{STOPPED}) from all \textit{daq\_reader} instances. Each \textit{daq\_reader} follows a complementary state sequence (\texttt{IDLE}, \texttt{INITIALIZING}, \texttt{INITIALIZED}, \texttt{STARTING}, \texttt{RUNNING}, \texttt{STOPPING}) in lockstep with these commands. This synchronous, state-gated protocol guarantees that all components proceed in coordinated phases, preventing race conditions and undefined behaviors during high-throughput operation.

\subsection{DAQ Reader}
\begin{figure}[htb]
    \centering
    \includegraphics[width=1.0\textwidth]{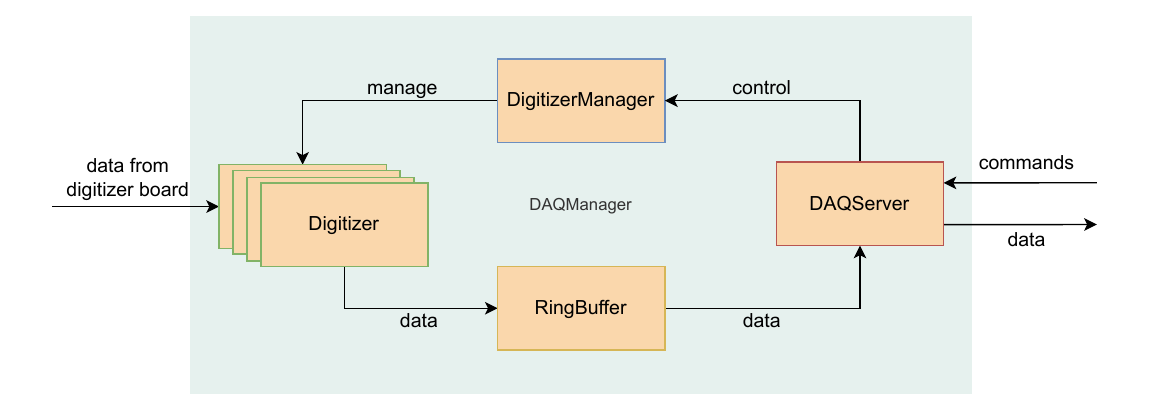}
    \caption{Architecture of the daq\_reader program.}
    \label{fig:DAQreader}
\end{figure}

The \textit{daq\_reader} program is responsible for receiving data from the digitizer boards, buffering it, and forwarding it unmodified to the \textit{collector} program. Its main architecture is illustrated in Fig.~\ref{fig:DAQreader}. The core component is the \textit{DAQManager}, which controls the entire application lifecycle. Several major components operate under its supervision, all implemented as custom C++ classes.

The \textit{Digitizer} class provides a software abstraction of a physical digitizer board. It integrates methods for configuration, start, and stop of data acquisition, and manages basic hardware parameters such as the digitizer name and IP address as member variables. In this system, a single DAQ server manages multiple physical digitizer boards. This is implemented in software by the \textit{DigitizerManager} class, which manages multiple instances of the \textit{Digitizer} class. The \textit{DigitizerManager} uniformly handles the lifecycle of all \textit{Digitizer} instances, including their initialization, start, and stop operations.

The \textit{DAQServer} class handles TCP communication. It creates a TCP server that listens for connections from the \textit{collector} program and manages them as separate sessions. The first connection from a remote host is designated as the \textit{control connection}, while subsequent connections are treated as \textit{data connections}. In the current implementation, four \textit{data connections} are established. The server is built on the Asio library, utilizing its asynchronous I/O model. All network operations are scheduled and executed by an \textit{io\_context} pool running on eight threads, enabling efficient concurrent handling of multiple sessions without dedicating a thread to each.

During a typical data acquisition run, control commands received by the \textit{DAQServer} via the control session are passed to the \textit{DigitizerManager}, which executes the corresponding operations. Specifically, upon receiving an \texttt{INIT} command, the \textit{DigitizerManager} initiates a mandatory configuration sequence. It first retrieves the complete, latest digitizer configuration in JSON format from a central PostgreSQL database. This global configuration contains setup parameters for all digitizer boards across the system. The \textit{daq\_reader} instance then filters this global configuration based on the \textit{hostname} of the server it is running on, extracting only the subset pertinent to its locally managed hardware. Using this host-specific configuration, the \textit{DigitizerManager} dynamically instantiates and initializes the corresponding \textit{Digitizer} objects. This design centralizes configuration management and ensures each server autonomously configures its assigned hardware components.

Following successful initialization, other control commands can be propagated to all managed digitizers. For data acquisition, the \textit{DigitizerManager} launches one acquisition thread for every digitizer. The acquired data blocks are temporarily stored in a ring buffer.

For data transmission, each data session posts asynchronous send operations. These operations are serviced by the aforementioned thread pool, which concurrently retrieves data blocks from the shared ring buffer and sends them to the \textit{collector}. The threads writing data from the digitizers into the ring buffer and the threads performing network I/O are fully decoupled. This design, based on an asynchronous producer-consumer pattern, simplifies lifecycle management and improves transmission stability and throughput. When acquisition is stopped, the \textit{DigitizerManager} terminates all acquisition threads. The \textit{DAQServer} ensures all pending asynchronous send operations complete, transmitting any remaining data in the ring buffer before shutting down.

\subsection{Collector}
\begin{figure}[htbp]
    \centering
    \includegraphics[width=1.0\textwidth]{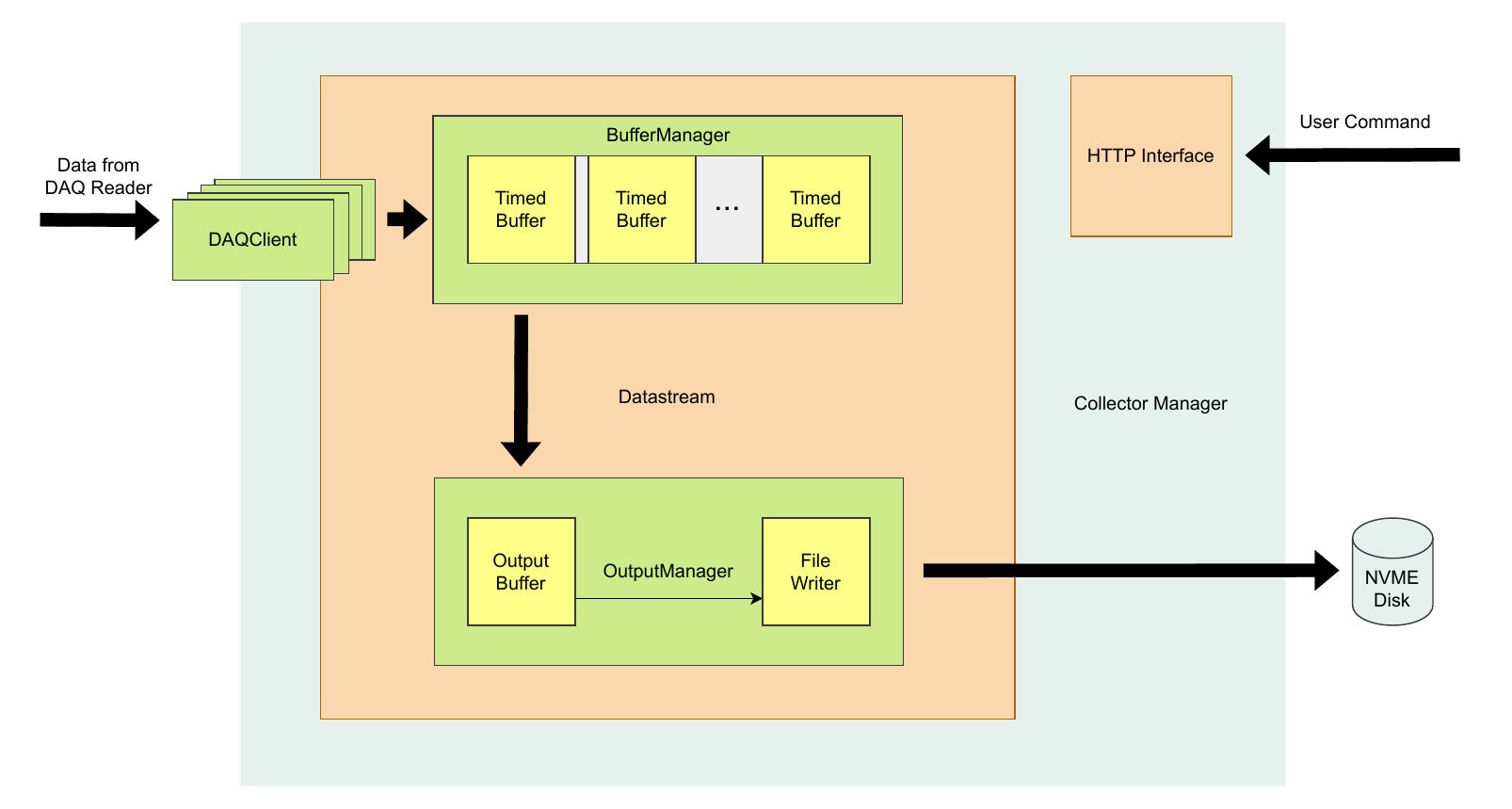}
    \caption{The architecture of the collector program.}
    \label{fig:collector}
\end{figure}

The \textit{collector} program receives data from multiple \textit{daq\_reader} instances, sorts them in memory, and writes the ordered data to disk. It also provides a user interface for controlling the entire data acquisition process. Its architecture is illustrated in Fig.~\ref{fig:collector}.

The program employs a \textit{CollectorManager} class to control its lifecycle. The main functional components under this manager are the \textit{DataStream}, \textit{BufferManager}, and \textit{OutputManager} classes. Incoming data are first organized into logical \textit{streams} based on the source IP addresses of the \textit{daq\_reader} instances; a single stream may aggregate data from multiple IPs. Each \textit{DataStream} is associated with its own \textit{BufferManager} and \textit{OutputManager}, forming an independent processing pipeline. This design allows for concurrent and scalable handling of data from different sources.

\subsubsection{Buffering and Sorting with the \textit{BufferManager}}
\begin{figure}[htb]
    \centering
    \includegraphics[width=1.0\textwidth]{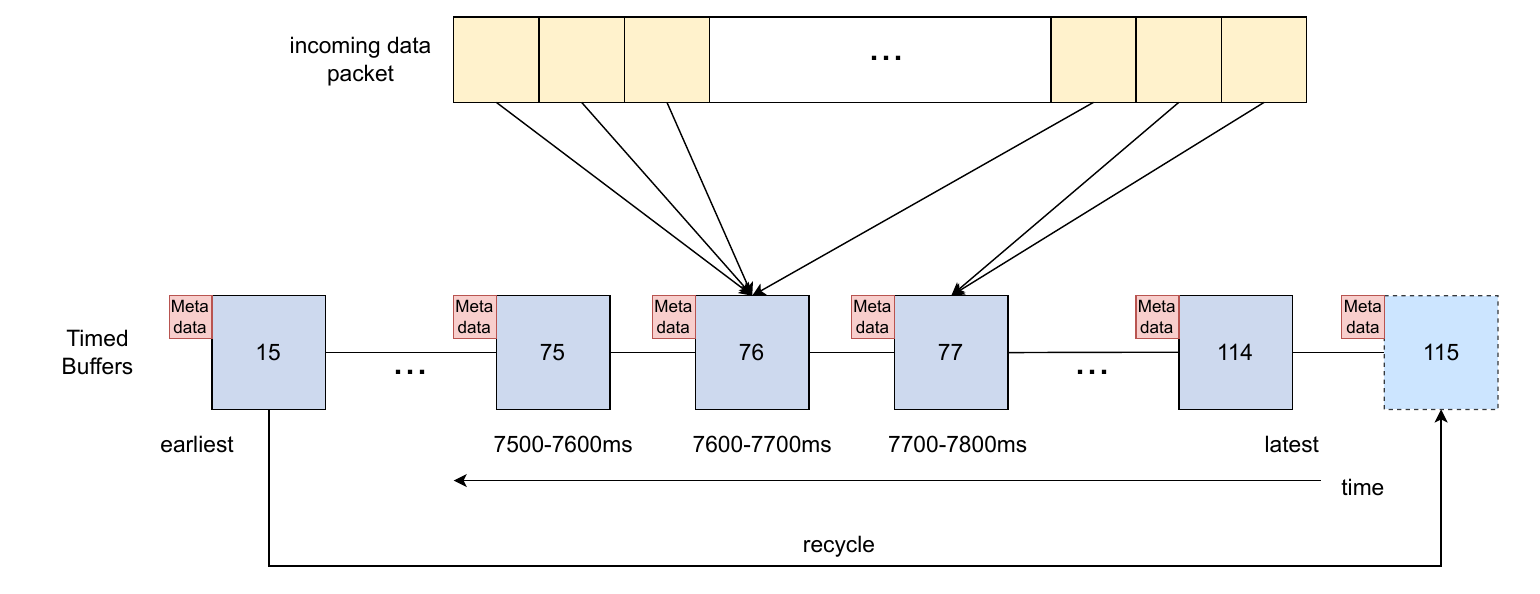}
    \caption{An example illustrating how data blocks from a packet are distributed into their corresponding timed buffers. The earliest timed buffer (leftmost) will be processed and then recycled to the end of the list for reuse.}
    \label{fig:BufferManager}
\end{figure}

The \textit{BufferManager} is responsible for receiving unsorted data blocks and performing an initial time-based organization. It manages a fixed-size list of \textit{timed buffers} (currently 100), each corresponding to a specific 100~ms time window and with a capacity of 512~MB. An example of how data blocks are distributed into their respective timed buffers is illustrated in Fig.~\ref{fig:BufferManager}.

Within this structure, each timed buffer is designated for a specific time interval, for example, holding data for the period 7500–7600~ms. When a data block arrives, its header is unpacked to extract the timestamp, channel number, and size. The raw data of the block is placed into the corresponding timed buffer based on its timestamp. Concurrently, a metadata object, containing the channel, timestamp, size, and memory location of the data block, is inserted into a dedicated container within the same timed buffer. This separation of data and lightweight metadata enables efficient subsequent sorting.

To accommodate network latency and potential delays in data arrival, the recycling of timed buffers does not begin until 6 seconds after the start of data acquisition. Thereafter, processing occurs in approximately 100~ms intervals, targeting the oldest timed buffer (e.g., the one covering the earliest 100~ms window after the initial 6-second mark).

The hardware trigger timestamp is used exclusively for ordering data blocks within each stream and is the sole criterion for the final on-disk order. The collector's system clock, however, serves a different role: it drives the software timers (\texttt{asio::steady\_timer}) that decide when a partially-filled buffer should be flushed. Because the two clocks are not perfectly synchronized, the system clock may drift relative to the hardware clock over long runs. If this drift becomes significant, the software-timer flushes may fire too early or too late; in either case an incoming data block may find no available buffer, stalling data taking. The correction described below compensates for this drift, ensuring consistent buffer management throughout a run.
After every 100 buffers processed (about every 10 seconds), the \textit{collector} retrieves the hardware clock information from the trigger board. It then calculates the time difference $\delta t$ between the hardware clock and the start time of the current first buffer, and compares this $\delta t$ with that obtained in the first cycle. If the difference (drift) is less than 1 ms, no adjustment is made; otherwise, a small adjustment of $\pm1$ or $\pm2$ ms is applied to the next waiting period. This ensures stable and accurate periodic processing despite clock drift.
Once a timed buffer is selected for processing, its metadata entries are first sorted chronologically. Then, based on this sorted order, the corresponding data blocks are read out and transferred to the \textit{OutputManager}. After processing, the timed buffer is cleared and recycled to the end of the list to store future data.

\subsubsection{Ordered Writing with the \textit{OutputManager}}
The \textit{OutputManager} receives the chronologically ordered data blocks from the \textit{BufferManager} and manages their final write to disk. 

It contains multiple output buffer slots. When a timed buffer is ready, the \textit{OutputManager} assigns an available output buffer slot. The data blocks, already in sorted order, are written contiguously into this slot. Each filled output buffer thus contains a complete, ordered 100~ms segment of data.

This extra copy deliberately trades a small CPU overhead for a large I/O efficiency gain. A pointer-based scheme would scatter the final disk write across many non-contiguous chunks of the timed buffer, producing a scatter-gather I/O pattern that cannot saturate modern NVMe bandwidth. By combining the small incoming blocks into one contiguous memory region, the second copy enables a single sequential \texttt{write()} system call that fully utilises the storage hardware.

A dedicated file-writing thread within the \textit{OutputManager} handles disk I/O. Once a timed buffer is fully dumped into an output buffer slot, it is marked as ready. The file-writing thread is notified, takes a snapshot of the buffer's pointer information (to ensure thread-safe access), and writes the data to disk. After the write completes, the output buffer slot is released back to the pool for reuse. This decoupled design, where the main thread fills buffers and a dedicated thread writes them, maximizes I/O efficiency.

File output in the \textit{collector} is organized chronologically. The time span of the output file, determined by the number of seconds, is a configurable parameter. During an acquisition run, the file index is incremented each time a file of the configured span is completed. The time span can also be updated dynamically while the \textit{collector} is in the \textbf{IDLE} state via its HTTP interface, allowing operators to adjust file granularity according to the bandwidth of different types of runs.

The naming format for these files is also configurable, with the sole requirement being that it incorporates the current \texttt{run} number and \texttt{file} index as variable parameters. A common convention is \path{run_%05d_file_%05d.data}, which would generate a filename like \path{run_00121_file_00000.data} for the first file of run 121.

When a data file is closed, its metadata, including the run number, file index, size in bytes, open and close time, and file path, is recorded in a PostgreSQL database. Simultaneously, a notification message containing the \texttt{run} and \texttt{file} information is published to a Kafka message queue. This message serves as a trigger for downstream automated processing pipelines, ensuring that new data files are promptly identified and processed for tasks such as online monitoring and data transfer. This integrated design enables a fully automated data handling workflow.

\subsubsection{Error Handling}
Data integrity is enforced at multiple levels within the \textit{collector}. Each incoming data block carries a protocol marker and a segment-size field that are validated before the block is accepted. The embedded timestamp is checked against the current buffer window; if it falls outside the expected range, the block is rejected, the error is reported as a log entry, and the system enters a safe state that halts data taking, preventing corrupted data from propagating to storage. All dynamically allocated memory is managed through RAII (Resource Acquisition Is Initialization) wrappers, guaranteeing that resources are reclaimed even on error paths. These mechanisms have been validated through unit tests with deliberately corrupted data, confirming that anomalies are correctly detected and the system transitions to a safe state without memory leaks.

\subsubsection{HTTP Control Interface}
The \textit{collector} provides a RESTful HTTP API~\cite{fielding2000rest} for operator control and monitoring, decoupled from its core data processing tasks. A dedicated thread hosts a lightweight HTTP server implemented using the \texttt{httplib} library~\cite{cpphttplib}. All control commands are issued by sending an HTTP POST request to the \textit{collector} server at the \textbf{/daq} endpoint.

The request body must be a JSON object containing at minimum a \texttt{cmd} key, which specifies the desired action. The value of \texttt{cmd} is a string from a defined set: \texttt{init}, \texttt{start}, \texttt{stop}, \texttt{status}, \texttt{reset}, \texttt{list config}, or \texttt{modify output duration}. Depending on the command, additional parameters may be included in the same JSON object. For example, a \texttt{modify output duration} includes the index of the target stream and the new file time span in seconds.

Upon receiving a valid command, the HTTP server thread forwards it to the main \textit{CollectorManager}. The manager executes the command, which may involve changing the internal state, propagating instructions to \textit{daq\_reader} instances via their control sessions, or updating runtime configuration. The server then returns a JSON-formatted response indicating success or failure, often accompanied by relevant status information. This design allows for seamless integration with external control scripts, graphical user interfaces, or automated run management systems.

\section{Performance Evaluation}
\label{sec:performance}
Due to the distributed architecture, the main load for data aggregation and processing is concentrated on the \textit{collector}. To locate potential bottlenecks, the internal data flow of the collector was broken down into key stages: network reception, copying data into the timed buffers (the \textit{copydata} stage), transferring sorted data from timed buffers to output slots (the \textit{bufferreader} stage), and final disk writing. The performance of the critical memory-copy stages (\textit{copydata} and \textit{bufferreader}) was measured using the Google Benchmark framework~\cite{googlebenchmark}, while network bandwidth and disk I/O capability were evaluated based on hardware specifications and standard system tools.

\subsection{Throughput Benchmarks}
The \textit{collector} receives data via multiple 10~Gbps SFP\texttt{+} links, providing an aggregate network bandwidth well in excess of the 1.6~GB/s requirement. Therefore, the internal memory-intensive processing stages were the primary focus of benchmarking. All benchmarks were conducted on a dedicated Dell R7625 server equipped with two AMD EPYC 9554 processors, 192 GB of DDR5 memory, and a Micron 7500 Pro NVMe SSD (3.84 TB). This platform ensures that performance measurements reflect the capabilities of the software logic rather than being limited by I/O or memory constraints.

\textbf{Copy to Timed Buffer:} The \textit{copydata} stage, executed by the Asio thread pool, unpacks incoming blocks and copies them into the appropriate timed buffer. Its performance was measured using the Google Benchmark library. Prepared data blocks with timestamps randomly distributed across a 100~ms window were copied. As shown in Table~\ref{tab:copydata_benchmark}, this stage achieves bandwidths significantly above 20~GB/s, confirming it is not a bottleneck.

\begin{table}[htbp]
\centering
\begin{tabular}{ccc}
\hline
Data Size (MB) & Time (ms) & Projected Bandwidth (GB/s) \\
\hline
100 & 6.71 & 14.9 \\
200 & 12.0 & 16.7 \\
300 & 11.7 & 25.6 \\
400 & 17.8 & 22.5 \\
500 & 23.7 & 21.1 \\
\hline
\end{tabular}
\caption{Benchmark results for the \textit{copydata} stage (copy into timed buffer).}
\label{tab:copydata_benchmark}
\end{table}

\textbf{Copy to Output Buffer Slot:} The \textit{bufferreader} stage sequentially copies data from a processed timed buffer into an output buffer slot. Benchmarks (Table~\ref{tab:bufferreader_benchmark}) show it sustains a bandwidth exceeding 3~GB/s, meeting the system target.

\begin{table}[htbp]
\centering
\begin{tabular}{ccc}
\hline
Data Size (MB) & Time (ms) & Projected Bandwidth (GB/s) \\
\hline
100 & 31.9 & 3.13 \\
200 & 54.8 & 3.64 \\
300 & 86.3 & 3.47 \\
400 & 121 & 3.30 \\
500 & 147 & 3.40 \\
\hline
\end{tabular}
\caption{Benchmark results for the \textit{bufferreader} stage (copy to output buffer slot).}
\label{tab:bufferreader_benchmark}
\end{table}

\textbf{Disk Write Performance:} The ultimate storage bottleneck was assessed using \textit{fio}. The storage subsystem, formatted with XFS and with proper partition alignment, demonstrated a stable write bandwidth of ~5.47~GB/s, which is sufficient for the target data rate.

The benchmark results confirm that all critical internal stages and the final disk I/O can sustain data rates exceeding 3~GB/s, which is far beyond the design target of 1.6 GB/s. The \textit{bufferreader} stage is identified as the limiting factor within the collector software, aligning closely with the system's maximum throughput goal.

In addition to these micro-benchmarks, system-level tests were conducted on the PandaX-20T prototype platform at the Tsung-Dao Lee Institute. With all 28 digitizer links active in self-trigger mode, the \textit{daq\_reader} sustained an aggregate throughput of approximately 1~GB/s while consuming roughly 12.6 CPU cores, which represents a small fraction of the available computing resources on the test server.  The \textit{collector} was tested under various trigger rates and achieved a peak throughput of approximately 2.8~GB/s, consistent with the 3~GB/s estimate from the micro-benchmarks and well above the 1.6~GB/s design target. These results confirm that the framework can meet its design throughput target with considerable headroom under realistic operating conditions.

\subsection{Stability Test}
Beyond throughput, long-term operational stability is essential. On the proto test platform for PandaX-20T equipped with more than 1400 channels, the system underwent a continuous data acquisition run for over 58 hours without encountering any software errors or data corruption.
This successful test surpasses the initial design goal of stable 24-hour operation, demonstrating the robustness of the framework under prolonged load.

\subsection{Validation in PandaX-4T data taking}
To validate the system under realistic experimental conditions, we deployed AURORA during the high-rate calibration campaign conducted at the conclusion of PandaX-4T operation. During this period, the detector was exposed to intense calibration sources, including an AmC source, a DD and DT generator, and injected $^{220}$Rn, generating sustained data rates that approached the design limits of the DAQ system.

During two AmC calibration runs lasting 11 and 13 hours respectively, the system sustained an average data rate of 800~MB/s. In several $^{220}$Rn injection and decay campaigns, each spanning approximately one full day (with about one-third of the time dedicated to injection and the remaining two-thirds to decay), peak data rates reached 900~MB/s and the average data rate exceeded 450~MB/s. Throughout these operations, the system performed reliably without any interruptions, confirming its capability to handle sustained high-rate data streams in a real experimental environment.

\section{Scalability and Adaptability}
\label{sec:scalability}

The design of the DAQ framework prioritizes not only performance and stability but also future extensibility and adaptation to diverse experimental needs. Its modular architecture allows for functional extensions and scaling to accommodate higher data rates or new physics objectives.

\subsection{Online Processing and Alert Generation}
The current \textit{OutputManager} primarily writes time-ordered data to persistent storage. Its architecture, however, inherently supports serving as a real-time data source. The sorted data blocks, residing in the output buffer slots before being written to disk, can be simultaneously streamed to auxiliary online analysis processes. This capability enables real-time monitoring, online quality checks, and even the implementation of low-latency trigger algorithms for rare event detection, such as core collapse supernova neutrino bursts or other transient phenomena. Adding such functionality would require minimal modification, primarily involving the establishment of additional inter-process communication channels to broadcast the buffer snapshots.

\subsection{Horizontal Scaling for Higher Channel Counts}
The present \textit{collector} instance is designed to handle the anticipated peak data rate. Should future detector upgrades lead to a total bandwidth exceeding the processing capacity of a single node, the architecture naturally supports a multi-\textit{collector} deployment. In this scenario, data streams from different subsets of \textit{daq\_reader} instances would be directed to separate, independently running \textit{collector} servers. Each server would produce its own set of chronologically organized data files. Since all files are strictly indexed by run number, file index, and inherently by timestamp, the data from parallel \textit{collectors} remain globally aligned in time. Subsequent offline analysis can seamlessly merge and process these files based on their timestamps, ensuring no compromise in data integrity or analysis capability.

\subsection{Generalization Beyond the PandaX Experiment}
While this framework was developed for the PandaX experiment, its core design is applicable to other high-throughput physical experiments. The fundamental requirement for adaptation is that the raw data format includes a reliable timestamp for each event or data block. The \textit{copydata} module, responsible for parsing the raw data header and extracting timing information, is the primary component that would require experiment-specific modification. The subsequent buffering, sorting, and output mechanisms, the \textit{BufferManager}, \textit{OutputManager}, and network layers, are largely agnostic to the specific data payload and can be reused with minimal change. This makes the framework a versatile foundation for building DAQ systems where high data rates and reliable, ordered storage are paramount.

\section{Summary}
\label{sec:summary}

This work presents the design and implementation of AURORA, a high-performance, distributed data acquisition (DAQ) framework developed for the next-generation PandaX dark matter experiment. Designed to meet the challenges posed by the PandaX-xT detector, including a high channel count and sustained multi-GB/s data rates, the system prioritizes data integrity, throughput stability, and long-term operational reliability.

The framework employs a distributed architecture, separating data reception and aggregation. Multiple \textit{daq\_reader} instances acquire data in parallel from digitizers and stream it via high-speed TCP connections to a centralized \textit{collector}. A key innovation is the multi-stage buffering and deferred-sorting strategy within the collector, which decouples data reception, time-ordered reorganization in memory, and disk I/O. This design effectively mitigates back pressure and ensures smooth data flow under high load. Benchmark results confirm that the critical processing stages exceed the design target of 1.6~GB/s.

The system has been successfully deployed on the dedicated proto test platform for PandaX-20T, operating stably with 176 digitizers at an aggregate throughput of 1.4~GB/s. A continuous 58-hour acquisition run without errors validated its robustness, exceeding the 24-hour design goal. Integrated system services, a web-based control interface, and real-time monitoring provide a complete and operator-friendly environment for sustained data-taking campaigns.

The architecture is fundamentally scalable and adaptable. It can horizontally scale to accommodate higher channel counts through multiple \textit{collector} instances, and its sorted data stream can be extended to feed online analysis and alert systems. While developed for PandaX, the core framework is readily adaptable to other high-throughput experiments where data blocks carry timestamps, requiring only minimal modifications to the data parsing layer. This work provides a proven, high-performance DAQ solution that meets the demands of upcoming PandaX runs and can serve as a reference design for future large-scale experiments in particle and nuclear physics.

\section*{Acknowledgments}
We thank Mr. Yuan Li for the helpful discussion on the original DAQ system and related electronics.
This project is supported by the grants from National Science Foundation
of China (No. 12175139).

\bibliographystyle{elsarticle-num} 
\bibliography{refs} 

\end{document}